\documentclass[published,notoc]{JHEP} 

\usepackage{epsfig}
\usepackage{amsmath}
\usepackage{curves}

\def\PT{P_\tau}
\def\MQ{\frac{m_\tau^2}{Q^2}}
\def\QM{\frac{Q^2}{m_\tau^2}}
\def\GN{(|g_V|^2+|g_A|^2)}
\def\KK{(|\kappa|^2+|\tilde{\kappa}|^2)}
\def\GVA{\gamma_{VA}}
\def\GK{\gamma_\kappa}
\def\TA1{\tau\rightarrow a_1\nu_\tau}
\def\TRO{\tau\rightarrow \rho\nu_\tau}
\def\T3M{\tau\rightarrow 3h\nu_\tau}
\def\TPI{\tau\rightarrow \pi\nu_\tau}
\def\TP2{\tau^-\rightarrow \pi^-\pi^0\nu_\tau}

\newcommand{\epj}[3]{{\it Europ. Phys. Journ. }{\bf C #1} (#2) #3}

\title{\bf Weak dipole moments in $\tau$ decays to hadrons}

\author{Pablo E. Lacentre \\
        Universidad Nacional de La Plata, cc67, 1900 La Plata, Argentina \\
        E-mail: \email{lacentre@venus.fisica.unlp.edu.ar} }

\received{\today}

\abstract{
A description of three body tau decays with general 
vectorial and derivative (non-minimal) couplings is presented for 
experiments where the tau direction of flight is not reconstructed. 
The proposed interactions may be due to charged current 
weak dipole moment couplings and hence to structure of the $\tau$ 
lepton. The decay distribution is expressed in terms of seven real 
parameters plus an overall normalization factor. Comments on the extension 
of the formalism to two-body decays are included. 
The particular cases of $\TA1$ and $\TP2$ are discussed in detail. 
}

\begin{document}


Data on hadronic $\tau$ decays provide an interesting window to
the search of new physics. This is particularly important for 
final states with higher pion multiplicity, which in some cases 
present inconsistency between theory and experiment \cite{Eidelman}.
Then, with the high precision measurements at LEP and SLD, tau physics 
offers us the possibility to search for non-Standard Model couplings.
One possibility is the proposal of derivative (non-minimal) couplings 
for the tau. The existence of these type of couplings would be
evidence of a tau being a composite object rather than a third generation 
lepton. This fact should manifest at the tau production vertex
\cite{Bernabeu,OPAL} and at the decay vertex \cite{Rizzo,Chizhov} 
with approximatly the same strength.
Nevertheless, the structure of both vertices can be explored 
independently if one does not deal with correlations betwen 
$\tau^+\tau^-$ pairs. 

In this paper the posibility of 
measuring derivative couplings in tau decays to hadrons is investigated.
A detailed calculation for events with three pseudoscalar
mesons in the final state is presented, but the formalism can be easily 
applied to decays into two mesons and even to decays with higher pion 
multiplicity. The notation of reference \cite{Kuhn} is followed throughout. 

The paper is organized as follows: after the introduction of the 
effective currents, the complete angular distribution is presented. 
This distribution is expressed in terms of seven real
parameters plus an overall normalization factor. Section three 
is devoted to the analysis of the total decay rate, the distribution 
of the tau decay into three pions and some comments on the two
pion  decay mode. Particular distributions involving strange 
particles will be presented in forthcoming papers.


The matrix element for the semileptonic tau decay into three 
(pseudo) scalar mesons is described in terms of hadronic ($H^\mu$) 
and leptonic ($J^\mu$) currents as
\begin{equation}
{\cal M} = \frac{G_F}{\sqrt{2}} \binom{\cos\theta_c}{\sin\theta_c} 
J^\mu H_\mu \;,
\label{E1}
\end{equation}
where $G_F$ is the Fermi constant and $\theta_c$ is the Cabibbo angle,
whose cosine (sine) has to be used in the case of final states with 
even (odd) number of kaons, implying $\Delta S=0$ ($\Delta S=1$). 

The leptonic current incorporates, in addition to the vector and axial-vector
structure, the weak charged current dipole moment couplings as
\begin{equation}
J^\mu=\bar{u}_\nu \left\{ \gamma^\mu (g_V-g_A\gamma^5)
           - \frac{i}{2 m_\tau} \sigma^{\mu\nu} Q_\nu (\kappa-
          i \tilde{\kappa}\gamma^5) \right\} u_\tau  \;,
\end{equation}
where $\sigma^{\mu\nu}=i/2 [ \gamma^\mu , \gamma^\nu] $,
$Q^\mu$ is the transfered momentum and the parameters $\kappa$ and 
$\tilde{\kappa}$ are the (CP-conserving) magnetic and (CP-violating) 
electric dipole form factors respectively. 
These parameters are in general complex quantities, but CPT 
conservation enforces that the imaginary part of $\tilde{\kappa}$ 
vanishes. Nevertheless, the analysis followed in this paper is not 
restricted to any particular case. The SM is constructed under the 
hypothesis $g_V=g_A=1$ and $\kappa=\tilde{\kappa}=0$. 

The most general ansatz for the hadronic current is characterized by four form factors:
\begin{equation}
H^\mu = V_1^\mu F_1 + V_2^\mu F_2 + i V_3^\mu F_3 + V_4^\mu F_4 \;,
\end{equation}
with
\begin{align}
V_1^\mu & = (q_1 - q_3)_\nu  T^{\mu\nu} \;, \nonumber \\ 
V_2^\mu & = (q_2 - q_3)_\nu  T^{\mu\nu} \;, \nonumber \\ 
V_1^\mu & = \epsilon^\mu_{\alpha\beta\gamma} q_{1\alpha} q_{2\beta} q_{3\gamma}  
\;, \nonumber \\ 
V_1^\mu & = q_1^\mu + q_2^\mu + q_3^\mu \equiv Q^\mu \;,
\end{align}
where
\begin{equation}
 T^{\mu\nu} = g^{\mu\nu} - \frac{Q^\mu Q^\nu}{Q^2}
\end{equation}
and $q_i$ are the pions 4-momenta.
Since the strong interaction conserves parity, the form factors $F_1$
and $F_2$ ($F_3$) originate from the axial (vector) hadronic current 
and correspond to a spin one state, while $F_4$ is due to the spin 
zero part of the axial current.
Specific models for various three meson final states based on chiral
symmetry were derived in \cite{Santamaria,Decker}. 

The differential decay rate is obtained via the calculation of the 
leptonic and hadronic tensors: $L^{\mu\nu}=J^\mu(J^\nu)^\dag$ and 
$W^{\mu\nu}=H^\mu(H^\nu)^\dag$.
These second rank tensors are conveniently decomposed in 16 symmetric and 
antisymmetric real combinations \cite{Kuhn} and their
contraction takes the form $L^{\mu\nu} W_{\mu\nu} = \GN (m_\tau^2-Q^2) 
\sum_{X=1}^{16} L_X W_X$.
The kinematical observables are defined in
the hadronic rest frame, $\vec{q_1}+\vec{q_2}+\vec{q_3}=0$, with the 
z-axis in the direction  of the laboratory system and the x-axis in the 
direction $\vec{q}_3$ of the meson with different charge 
(see figure \ref{F1}): 

\begin{itemize}
\item the invariant hadronic mass squared $Q^2$ and 
\begin{equation}
s_i \equiv (q_j+q_k)^2\;,  \qquad  i,j,k=1,2,3 \;, \quad i\neq j\neq k  \;;
\end{equation}
\item the direction of the normal to the decay plane%
\footnote{see \cite{Privitera} for expressions in terms of laboratory observables.}
$\vec{n_\perp} \equiv \vec{q}_1 \times \vec{q}_2$: 
\begin{align}
\cos\beta  & \equiv \hat{n}_L . \hat{n}_\perp   \;, \\
\cos\gamma & \equiv -\frac{\hat{n}_L . \vec{q}_3}{ | \hat{n}_L 
             \times \hat{n}_\perp | }   
             \;, \nonumber
\end{align}
\item the angle between the tau spin-vector ($\vec{s}$) and the 
hadronic direction in the $\tau$ rest frame:
\begin{equation}
\cos\theta = \frac{2xm_\tau^2-m_\tau^2-Q^2}{(m_\tau^2-Q^2)
             \sqrt{1-4m_\tau^2/s}} \;,
\end{equation}
with
\begin{equation}
x=\frac{2E_h}{\sqrt{s}} \;, \qquad s=4E_{beam}^2
\end{equation}
\end{itemize}

After integration over the unobservable neutrino direction and over 
the third Euler angle $\alpha$ the differential decay rate is given 
by%
\footnote{$\alpha$ is observable only in the case where the $\tau$ direction 
is measured.}
\begin{align} \label{E3}
d\Gamma & = \frac{G_F^2}{4 m_\tau} \, \GN \, \binom{\cos\theta_c}
          {\sin\theta_c}
          \left\{ \sum_X L_X W_X \right\} \\
\mbox{} & \times \frac{1}{(2\pi)^5} \frac{1}{64} \, \frac{(m_\tau^2
          -Q^2)^2}{m_\tau^2} 
          \frac{dQ^2}{Q^2} ds_1 ds_2 \frac{d\cos\theta}{2}
          \frac{d\cos\beta}{2} \frac{d\gamma}{2\pi}  \;. \nonumber
\end{align}
The leptonic functions $L_x$ take the form:
\begin{align} \label{Leptonic}
L_A & =   \frac{2}{3} K_1+K_2+K_6+\frac{1}{3} \bar{K}_1 
        \frac{3\cos^2\beta-1}{2}  \;, \nonumber \\
L_B & =   \frac{2}{3} K_1+K_2+K_6-\frac{2}{3} \bar{K}_1 
        \frac{3\cos^2\beta-1}{2}  \;,  \nonumber \\
L_C & =   -\frac{1}{2} \bar{K}_1 \cos{2\gamma} \, \sin^2\beta  \;,  \nonumber \\
L_D & =   \frac{1}{2} \bar{K}_1 \sin{2\gamma} \, \sin^2\beta  \;,  \nonumber \\
L_E & =   \bar{K}_3 \cos\beta  \;,  \nonumber \\
L_F & =   \frac{1}{2} \bar{K}_1 \sin{2\beta} \, \cos\gamma  \;,   \nonumber \\
L_G & =   -\bar{K}_3 \sin\beta  \, \sin\gamma  \;,   \nonumber \\
L_H & =   -\frac{1}{2} \bar{K}_1 \sin{2\beta} \, \sin\gamma  \;,    \nonumber \\
L_I & =   -\bar{K}_3 \sin\beta \, \cos\gamma   \;,  \nonumber \\
L_{SA} & =   K_2   \;,  \nonumber \\
L_{SB} & =   \bar{K}_2 \sin\beta \, \cos\gamma   \;,  \nonumber \\
L_{SC} & =   \bar{K}_7 \sin\beta \, \cos\gamma   \;,  \nonumber \\
L_{SD} & =   -\bar{K}_2 \sin\beta \, \sin\gamma  \;,   \nonumber \\
L_{SE} & =   -\bar{K}_7 \sin\beta \, \sin\gamma   \;,  \nonumber \\
L_{SF} & =   -\bar{K}_2 \cos\beta   \;, \nonumber \\
L_{SG} & =   -\bar{K}_7 \cos\beta  \;, 
\end{align}
where
\begin{align}
K_1     & =   1-\MQ-(1+\MQ) \GVA \PT \cos\theta  \nonumber \\
\mbox{} & + (1-\QM) \zeta +(1+\QM) \GK \PT \cos\theta
          - \beta \PT \cos\theta  \;,  \nonumber \\
K_2 & =   \MQ (1+\GVA \PT \cos\theta)  \;,   \nonumber \\
K_3 & =   \GVA-\PT \cos\theta - (\GK+\zeta \PT \cos\theta)
      + \frac{1}{2} (\beta-\alpha \PT \cos\theta)   \;,  \nonumber \\
K_4 & =   \sqrt{\MQ} \left[ \GVA+\frac{1}{4}\beta
      + \QM (\frac{1}{4}\beta-\GK) \right] \PT \sin\theta \;, \nonumber \\
K_5 & =   \sqrt{\MQ} \left[ 1+2\zeta+\frac{3}{4} \alpha
      - \QM (\zeta+\frac{1}{4}\alpha) \right] \PT \sin\theta \;, \nonumber \\
K_6 & =   \frac{1}{2} (\alpha+\beta \PT \cos\theta)
      + \QM (\zeta-\GK \PT \cos\theta)  \;, \nonumber \\
K_7 & =   \frac{1}{4} (\alpha'+\beta' \PT \cos\theta)  \;, \nonumber \\
K_8 & =   \frac{1}{4} \frac{m_\tau}{\sqrt{Q^2}} \beta' \PT \sin\theta  \;, 
          \nonumber \\
\bar{K}_1 & =   K_1 \frac{3\cos^2\psi-1}{2}
            - \frac{3}{2} K_4 \sin{2\psi}  \;, \nonumber \\
\bar{K}_2 & =   \tilde{K}_2 \cos\psi + \tilde{K}_4 \sin\psi  \;, \nonumber \\
\bar{K}_3 & =   K_3 \cos\psi - K_5 \sin\psi  \;, \nonumber \\
\tilde{K}_2 & =   K_2 + \frac{1}{4} (\alpha+\beta \PT \cos\theta)  
                \;, \nonumber \\
\tilde{K}_4 & =   K_4 - \sqrt{\QM} (\frac{\beta}{4}-\GK) 
                \PT \sin\theta)  \;,  \nonumber \\
\bar{K}_7 & =   K_7 \cos\psi + K_8 \sin\psi  \;, 
\end{align}
and%
\footnote{ultra-relativistic approximation valid ie. at LEP.}
\begin{xalignat}{2}
\cos\psi & = \frac{\eta + \cos\theta}{1+\eta \cos\theta} \;, & \qquad
\eta     & = \frac{m_\tau^2-Q^2}{m_\tau^2+Q^2} \;,
\end{xalignat}
with
\begin{align}
\GVA & = \frac{2 \text{Re}(g_v g_A^*)}{\GN}  \;, \nonumber \\
\GK  & = \frac{2 \text{Im}(\kappa \tilde{\kappa}^*)}{\GN}  \;, \nonumber \\
\alpha & = \frac{2\text{Re}(g_v \kappa^*) - 2\text{Im}(g_A \tilde{\kappa}^*)}{\GN}  
           \;, \nonumber \\
\beta  & = \frac{2\text{Re}(g_A \kappa^*) - 2\text{Im}(g_V \tilde{\kappa}^*)}{\GN}  
           \;, \nonumber \\
\alpha' & = \frac{2\text{Im}(g_v \kappa^*) + 2\text{Re}(g_A \tilde{\kappa}^*)}{\GN}  
           \;, \nonumber \\
\beta'  & = \frac{2\text{Im}(g_A \kappa^*) + 2\text{Re}(g_V \tilde{\kappa}^*)}{\GN}  
           \;, \nonumber \\
\zeta  & = \frac{1}{4} \frac{\KK}{\GN} \;.
\label{Parameters}
\end{align}
The functions $W_x$ expressed in terms of $Q^2$, $s_1$ 
and $s_2$ can be found in \cite{Kuhn}.
This is the most complete description of $\T3M$ decays involving 
derivative couplings.


\FIGURE[ht]{
  \setlength{\unitlength}{6.4mm}
  \begin{picture}(13,10)(2,0) 
    \put(4,4){\vector(0,1){4}}
    \put(4.2,8){\makebox(0,0)[lt]{$\vec{z}=\hat{n}_\perp$}}
    \put(4,4){\vector(-2,3){1.5}}
    \put(2.3,6.3){\makebox(0,0)[b]{${\hat{n}_{\text{lab}}}$}}
    \put(4,4){\vector(4,-1){2.65}}
    \put(6.8,3){\makebox(0,0)[tc]{${\hat{n}_\tau}$}}
    \curve(4,5.9, 3.5,5.8, 3.1,5.5) 
    \put(3.4,5.9){\makebox(0,0)[b]{${\beta}$}}
    \curve(3.7,4.6, 4.6,4.5, 5,3.7)
    \put(4.7,4.6){\makebox(0,0)[lb]{$\psi$}}
    \put(4,4){\circle*{.2}}
    \put(4.2,3.7){\makebox(0,0)[lt]{$h$}}
    \put(12,2){\vector(-4,1){2.6}}
    \put(9.8,2.6){\makebox(0,0)[lb]{${ \hat{n}_{h}}$}}
    \put(12,2){\vector(4,-1){2.6}}
    \put(14.6,1.3){\makebox(0,0)[lb]{${ \hat{n}_{\nu_\tau}}$}}
    \put(12,2){\vector(1,1){1.8}}
    \put(13.9,3.8){\makebox(0,0)[lt]{${ \hat{s}_\tau}$}}
    \put(12,2){\vector(-1,-1){1.8}}
    \put(10.6,0.3){\makebox(0,0)[lc]{${ \hat{n}_{\text{lab}}}$}}
    \curve(11.3,2.2, 11.9,2.4, 12.4,2.4) 
    \put(11.8,2.5){\makebox(0,0)[lb]{${\theta}$}}
    \put(12,2){\circle*{.2}}
    \put(12,1.7){\makebox(0,0)[t]{${ \tau}$}}
    \curvedashes[1mm]{0,1,1}
    \curve(6.7,3.3, 9.5,2.6)
    \curvedashes{}
  \end{picture} \label{F1}
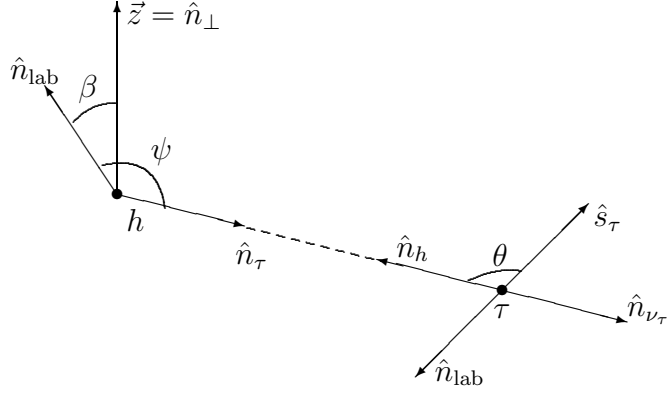
\caption{\em Observable angles
in the hadronic and $\tau$ rest frames.}
}



Let us consider now the modification of some experimental observables with
respect to the Standard Model predictions.
\begin{itemize}
\item The {\bf total rate} is given by:
\begin{align}
\Gamma  & =  \frac{G_F}{4 m_\tau} \frac{\GN}{4\pi} \cos^2\theta_c  \\ 
\mbox{} & \times \int dQ^2 (m_\tau^2-Q^2)^2 \left\{ \rho_0 + 
(1+\eta\frac{2 Q^2}{m_\tau^2}) \rho_1 \right\} \;,  \nonumber
\end{align}
where
\begin{equation}
\eta = 1+\frac{3}{4}\alpha+(1+\frac{Q^2}{2 m_\tau^2})
\zeta  \;,
\end{equation}
and $\rho_0$ and $\rho_1$ are the spin zero and spin one spectral 
functions, related to the structure functions $W_{SA}$ and $W_A+W_B$ 
respectively:
\begin{align}
\rho_0 & = \frac{1}{(4\pi)^4} \frac{1}{2 Q^4} \int ds_1 ds_2 W_{SA} 
           \;, \nonumber\\
\rho_1 & = \frac{1}{(4\pi)^4} \frac{1}{6 Q^4} \int ds_1 ds_2 (W_A + 
           W_B)  \;.
\end{align}
Notice the modification with respect to the SM by an enhancement 
factor $\eta$ in the spin one part, while the spin zero part is not 
affected. As a consecuence, the $\TPI$ decay is not sensitive to weak 
dipole moment couplings.
\item {\large $\mathbf{\TA1}$}:
The Cabbibo-allowed three pion decay of the $\tau$ proceeds 
dominantly via a state of $J^P=1^+$ denoted by $a_1(1260)$.
This is one of the most likely decay channels of the $\tau$, 
with a total branching ratio (3-prong plus 1-prong modes) of 18.71\% 
\cite{PDG}. Due to symmetry constraints, only the structure functions 
$W_A$, $W_C$, $W_D$ and $W_E$ are different from zero. 
The sensitivity of this channel to the parameters
can be computed following the ideas of reference \cite{Franzini}:
for a parameter $p_i$ characterizing a distribution function $f$, 
we define the sensitivity to this parameter as
\begin{align}\label{E4}
	S & \equiv \frac{1}{\sigma\sqrt{N}}    \\
  \mbox{} & = \left[ \int d\text{PS} \frac{1}{f} \left( \frac{
  \partial f}{\partial p_i} \right)^2
              \right]^{1/2}  \qquad  d\text{PS: phase space,}  
           \nonumber
\end{align}
where $\sigma$ is the error on the parameter determination for a 
given number of events $N$.
The sensitivity of the $\TA1$ decay mode to the parameters defined 
in (\ref{Parameters}) is given in table \ref{T1} (for comparison 
with table 1 in reference \cite{Rouge}).
\item {\large $\mathbf{\TRO}$}:
The formalism described above may be applied to semileptonic tau 
decays into two mesons (see \cite{Kuhn} for details). For the decay 
$\TP2$, which proceeds dominantly via the $\rho(770)$ resonance, the 
only structure function different from zero is $W_B$.
A complete analysis has been carried out in \cite{Rho}.
I show here only the sensitivity of this channel to the parameters
determination. 
\end{itemize}
Notice that $\alpha'$ and $\beta'$ do not influence the
distributions of these decay modes, so the corresponding 
sensitivities are zero and they are not given in the table.
The value obtained for each parameter is computed fixing the 
others to their SM-values and $\PT=0$.


\TABLE[ht]{
\begin{tabular}{|l|c|c|c|c|c|} \hline
  Channel   &  $\GVA$  &  $\GK$  &  $\alpha$  &  $\beta$  &  $\zeta$ 
            \\  \hline \hline
  $\TA1$    &  0.103  &  0.092 &  0.349 & 0.047  & 0.582  \\ 
  $\TRO$    &  0.042  &  0.024 &  0.232 & 0.009  & 0.354  \\ \hline
\end{tabular} \label{T1}
\caption{\em Sensitivity of the $a_1$- and $\rho$-decay modes to the 
parameters defined in expression (\ref{Parameters}). }
}


In conclusion, the most complete description of $\T3M$ decays involving 
derivative couplings has been derived.
The formalism is also applied to two body decays, which together 
constitute more than 50\% of the $\tau$ decays. Without 
reconstruction of the tau direction of flight all the available 
information is contained in the leptonic functions (\ref{Leptonic}) 
with the sensitivity given in table \ref{T1}. Of particular interest 
are the functions $L_{SC,SE,SG}$, which are zero in the SM. 
Hence, they would provide a direct evidence of weak dipole moment 
couplings and also the posibility to test CP-violation effects 
in tau decays. \\

The author would like to thank L.N. Epele and M.T. Dova for carefully 
reading this manuscript.
This work has been supported by CONICET, Argentina.



\begin{thebibliography}{999}

\bibitem{Eidelman} S. Eidelman and V. Ivanchenko, {\it Workshop on Tau Lepton 
         Physics} ({\bf Santander, 1998}), SLAC-SPIRES Conference {\bf C 
        98/09/07}.
\bibitem{Bernabeu} J. Bernab\'eu {\em et al.}, \plb{326}{1994}{168}
\bibitem{OPAL} K. Ackerstaet {\em et al.}, OPAL Collab., \plb{431}{1998}{188}
        {\bf B 431} (1998) 188.
\bibitem{Rizzo}   T. Rizzo, \prd{56}{1997}{3074}
\bibitem{Chizhov} M. Chizhov, \hepph{9612399} (unpublished).
\bibitem{Kuhn} J.H. K\"uhn and E. Mirkes, \zpc{56}{1992}{661}.
\bibitem{Santamaria} J.H. K\"uhn and A. Santamaria, \zpc{48}{1990}{445}.
\bibitem{Decker} R. Decker {\em et al.}, \zpc{58}{1993}{445}.
\bibitem{Privitera} P. Privitera, \plb{308}{1993}{163}.
\bibitem{Finkemeier} M. Finkemeier and E. Mirkes, \zpc{69}{1996}{4403}.
\bibitem{PDG} C. Caso {\em et al.} (Particle Data Group), \epj{3}{1998}{1}.
\bibitem{Franzini} P. Franzini, {\it The Da$\Phi$ne Physics Handbook}, Vol. 1 
        (1992) Chap. 1.
\bibitem{Rouge} A. Rouge, \zpc{48}{1990}{75}.
\bibitem{Rho} M.T. Dova \hepph{9904242}. Submitted to jHEP (1999). 

\end{thebibliography}
\end{document}